\documentclass[prl,twocolumn,showpacs,floats,superscriptaddress]{revtex4}
\usepackage{graphicx}
\usepackage{amsmath}

\graphicspath{{images/eps/},{images/pdf/}}

\begin{document}

\title{Friction at Atomic-Scale Surface Steps: Experiment and Theory}

\author{Hendrik H\"olscher}
\email{hendrik.hoelscher@imt.fzk.de}
\affiliation{Institute for Microstructure Technology, Forschungszentrum
Karlsruhe, P.O. Box 3670, 76021 Karlsruhe, Germany}
\author{Daniel Ebeling}
\affiliation{Center for NanoTechnology (CeNTech), University of M\"unster,
Heisenbergstrasse 11, 48149 M\"unster, Germany}
\author{Udo D. Schwarz}
\affiliation{Department of Mechanical Engineering and Center for Research
on Interface Structures and Phenomena (CRISP), Yale University,
P.O.\ Box 208284, New Haven, CT 06520-8284, USA}

\date{\today}

\begin{abstract}
Experiments performed by friction force microscopy at atomic-scale surface 
steps on graphite, MoS$_2$, and NaCl in ambient conditions are presented. 
Both step-down and step-up scans exhibit higher frictional forces at the edge, 
but distinguish in their load dependence: While the additional frictional 
force due to the step edge increases linearly with load if the tip has to 
jump a step up, it remains constant for downward jumps. This phenomena 
represents a universal effect that can be explained in terms of a modified 
Prandtl-Tomlinson model featuring a Schoebel-Ehrlich barrier at steps.
\end{abstract}

\maketitle

Tribology -- the science of friction, wear, and lubrication -- has impact on
many fields of science and technology. Consequently, it has been the subject
of intense research during the last centuries \cite{Dowson1979}. With the
advent of new experimental techniques such as \emph{friction force
microscopy} (FFM) \cite{Mate1987}, the study of frictional phenomena at the
atomic scale became accessible to tribologists, and the field of
\emph{nanotribology} has been established since (see, e.g.,
Refs.\,\onlinecite{Bhushan1995a,Carpick1997b,Urbakh2004a,Hoelscher2008a}).

The basic paradigm of nanotribological research is that the frictional 
behavior of a single asperity contact needs to be clarified in order to better
understand friction in complex macroscopic systems. Naturally, this makes the
friction force microscope a tool of choice for nanotribology. So far, most
experimental FFM studies designed to elucidate the atomic-scale principles of
friction focused on the frictional behavior of atomically flat surfaces (see,
e.g., Refs.\,\onlinecite{Luethi1996a,Morita1996,Kerssemakers1995a,%
Hoelscher1998a,Hoelscher1999c,Socoliuc2004a,Medyanik2006}). The stick-slip
phenomena observed in these experiments can be understood in the framework of
the well established and surprisingly simple \emph{Prandtl-Tomlinson} (PT)
model \cite{Prandtl1928,Tomlinson1929}, which is sometimes also referred to as
independent oscillator model \cite{McClelland1989,Helman1994}. An extension of
this model, obtained by including thermal activation processes, helped to
understand the velocity and temperature dependence of friction
\cite{Gnecco2000,Riedo2003,Schirmeisen2005a}.

Although these studies enabled valuable insight into the origin of atomic-scale
friction, atomically flat surfaces represent a simplified model case, since
any truly advanced model must include roughness. From this point of view, it
is surprising that only very few studies \cite{GMeyer1990b,Ruan1994b,%
Weilandt1995,EMeyer1996a,Mueller1997} focusing on the analysis of friction at
atomic-scale surface steps exist \cite{footnote}. All of  them reported
increased frictional forces at step edges compared to the value found on
atomically flat terraces, which has potentially far-reaching implications on
the friction observed in macroscopic systems. Meyer \textit{et al.}
\cite{EMeyer1996a} attributed this effect to the influence of the
\emph{Schwoebel-Ehrlich barrier} \cite{Ehrlich1966b,Schwoebel1966} present at
atomic step edges. Analyzing the load dependence of an Si$_{3}$N$_{4}$ tip on
graphite in ultrahigh vacuum, M\"uller \textit{et al.} \cite{Mueller1997}
found a directional dependence: Despite being much larger compared to the
on-terrace value, frictional forces were load-independent when jumping a step
down (``downwards scan''). In contrast, they increased linearly with load if
the tip had to move a step up (``upwards scan'').

\begin{figure}[tb]
\begin{center}
\includegraphics[width=\linewidth]{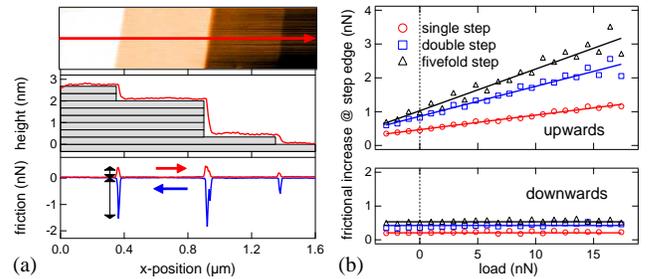}
\end{center}
\caption{(Color online) The frictional behavior of graphite at atomic-scale 
surface steps. \textbf{(a)} Top panel: Color-coded top view of the topography; 
image size: 1.6~$\mu$m $\times$ 0.4~$\mu$m. Middle panel: Topographical line 
section along the arrow indicated in the top panel. Individual graphene layers 
are represented by grey rectangles for illustration purposes, revealing that 
a double, a fivefold, and a single step are encountered. Bottom panel: 
Frictional forces recorded during a left-right scan (red) and the 
corresponding right-left scan (blue) along the same line with an external 
load of $F_{\rm load} = 6.8$\,nN. An higher frictional increase of the frictional
forces for upward than for downward scans is obvious (black arrows). 
\textbf{(b)} Plots of the frictional increase (defined as the difference 
between the maximum frictional force at the step edge and the friction 
encountered on the terrace) observed at the three different step edges as 
a function of the load. For upward scans (top panel), the frictional increase 
grows linearly with load, while it is constant for downward scans (bottom 
panel). Parameters: $c_z = 0.073$\,N/m, $c_x = 20.6$\,N/m, 
$v_{\rm M} = 6$\,$\mu$m/s.}
\label{fig:HOPG}
\end{figure}

The results presented in this letter demonstrate that the findings by M\"uller
\textit{et al.} are not restricted to graphite surfaces under vacuum
environment, but appear to be of general significance. Our investigations were
performed on freshly cleaved graphite(0001), MoS$_2$(001), and NaCl(001)
surfaces using rectangular silicon cantilevers (ContGD, BudgetSensors) in a
commercial friction force microscope (MultiMode AFM with Nanoscope IIIa
electronics by Veeco Instruments, Inc.) operated under ambient conditions. The
normal and lateral spring constants $c_{\rm z}$ and $c_{\rm x}$ as well as
friction and load were determined applying the calibration procedures
described in Ref.\,\onlinecite{Schwarz1996a}.

\begin{figure}[tb]
\includegraphics[width=0.9\linewidth]{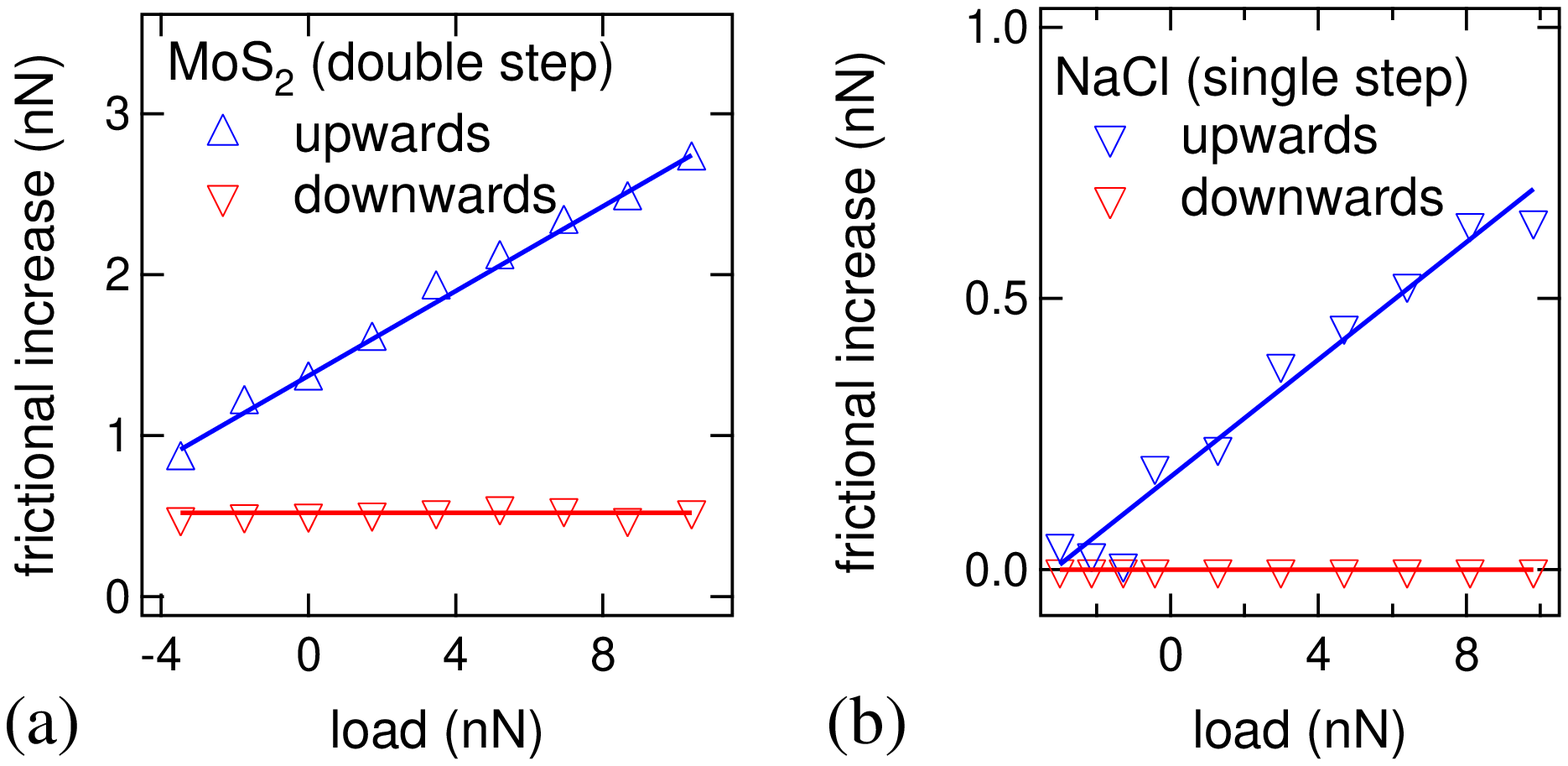}
\caption{(Color online) The frictional increase observed at a double step on 
MoS$_2$ \textbf{(a)} and a single step on NaCl \textbf{(b)}. On both materials,
the additional frictional forces caused by the step edges increase linearly 
with load for upward scans while they are independent of the actual loading 
force for downward scans. Note that in agreement with a previous report by 
Meyer and Amer \cite{GMeyer1990b}, we observe no change of friction at the 
edge compared to the on-terrace value for downward scans on NaCl, which leads 
to a vanishing value for the frictional increase. Parameters: 
$c_z = 0.065$\,N/m, $c_x = 18.2$\,N/m, $v_{\rm M} = 2$\,$\mu$m/s \textbf{(a)}, 
and $c_z = 0.063$\,N/m, $c_x = 17.8$\,N/m, $v_{\rm M} = 2$\,$\mu$m/s 
\textbf{(b)}.}
\label{fig:mos2}
\end{figure}

Figure \ref{fig:HOPG} summarizes the data obtained on graphite, where we
measured the frictional forces at three different step edges of one, two, and
five graphene layers height within the same scan. Maximum frictional forces
encountered at a step edge were always higher for upwards scans than for
downward scans [cf.\ Fig.\,\ref{fig:HOPG}(a), bottom panel, for illustration].
In order to adequately study the additional contribution of the step edge to 
the overall friction systematically as a function of the externally applied 
load $F_{\rm load}$, we always plot in the following the \textit{difference} 
between the frictional force needed to overcome the surface step and the 
frictional forces on the terraces, which we refer to as 
``frictional increase''. Our analysis of this additional, step edge-induced 
component revealed that it increases linearly for upward scans, while it 
remains constant for downward scans [Fig.\,\ref{fig:HOPG}(b)]. Note that we 
recover the same qualitative relationships for all three step heights, even 
though higher step heights show higher absolute increases.
Control experiments carried out on a double step on MoS$_2$ 
[Fig.\,\ref{fig:mos2}(a)] and a single step-edge on NaCl(001) 
[Fig.\,\ref{fig:mos2}(b)] reveal the same behavior.

Combining the previously published reports with our experimental findings
suggests that the described load dependence of friction at step edges is
likely to be a general phenomena and independent of both the sample material
as well as the specifics of the environment (vacuum or air). However, despite
this fairly far-reaching suspected range of validity and its consequential
impact on macroscopic friction, this effect has not been the subject of
thorough theoretical analysis yet. To fill this gap, we propose an extended
Prandtl-Tomlinson model that includes an explicit description of the
tip-sample interaction at atomic-scale surface steps and, as a result,
correctly reproduces the load dependence at atomic-scale surface steps.

\begin{figure}[bt]
\begin{center}
\includegraphics[width=0.55\linewidth]{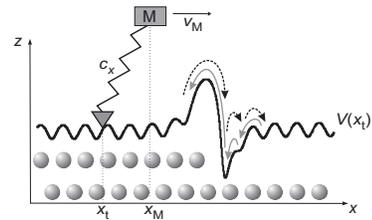}
\end{center}
\caption{A schematic of the modified Prandtl-Tomlinson model describing the 
friction at atomic-scale surface steps (not to scale). $x_{\rm t}$ represents 
the position of the tip, which is connected via a spring with spring constant 
$c_x$ to the body M. For sliding, the body M is moved along the $x$ direction 
while the tip interacts with the tip-sample potential $V(x_{\rm t})$ (thick 
solid line). If $x_{\rm t} = x_{\rm M}$, the spring is in its equilibrium 
position. The tip movement in the interaction potential at the step edge is 
indicated by arrows. For details, see text.}
\label{fig:PrandtlTomlinson}
\end{figure}

A schematic representation of the model is shown in
Fig.\,\ref{fig:PrandtlTomlinson}. A point-like tip is elastically coupled to a
main body $M$ with a spring possessing a spring constant $c_x$ in $x$ direction
and interacts with the sample surface via an interaction potential
$V_{\rm ts}(x_{\rm t},z_{\rm t})$, where $x_{\rm t}$ and $z_{\rm t}$ reflect
the actual position of the tip. The body $M$ experiences a constant loading
force $F_{\rm load}$ while it is scanned over the sample surface with a
velocity $v_{\rm M}$. The path of the tip can be calculated from the equations
of motion \cite{Hoelscher1997a}
\begin{subequations}
	\begin{align}
	m_{x} \ddot{x}_{\rm t} =  c_{x} (v_{\rm M}t - x_{\rm t})
		&-\frac{\partial~V_{\rm ts}}{\partial x_{\rm t}} -
		\gamma_{x} \dot{x}_{\rm t} \\
	m_{z} \ddot{z}_{\rm t} = -F_{\rm load}
		&-\frac{\partial~V_{\rm ts}}{\partial z_{\rm t}} -
		\gamma_{z} \dot{z}_{\rm t} \; ,
	\end{align}
\label{eq:DGL}
\end{subequations}
where $m_x$, $m_z$ are the effective masses of the system and $\gamma_x$,
$\gamma_z$ damping constants.

Within the PT model, the point-like tip represents the average of the actual
tip-sample contact, which might include several dozens or even hundreds of
atoms. The effect of the finite contact size is then mapped into the specific
choice of the interaction potential for calculation convenience. In the
classical PT model, this interaction is typically approximated by a simple
sinusoidal term. To include the effect of scanning over a step edge, we extend
this approach by explicitly introducing the tip-sample interaction at surface
steps into  $V_{\rm ts}(x_{\rm t},z_{\rm t})$. Due to the lack of a suitable
analytical description, we used a numerical approach as described below,
which ultimately allowed to recover all experimentally observed characteristics
from the simulations.

\begin{figure}[tb]
\begin{center}
\includegraphics[width=0.85\linewidth]{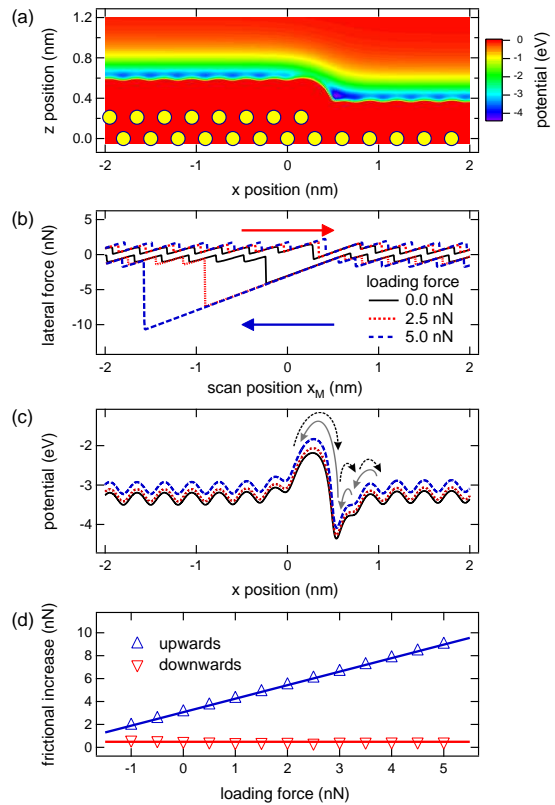}
\end{center}
\caption{(Color online)
\textbf{(a)} Density plot of the tip-sample interaction potential assumed in 
the simulations, featuring a barrier at the top of the step edge and a 
distinguished potential minimum at its bottom (dark blue). 
\textbf{(b)} The frictional force acting on the tip (offsets added for clarity)
exhibits a distinguished stick-slip movement at the step edge for down- and 
upward scans. Due to the relative increase of the barrier height and slope 
with load, the lateral force needed to pull the tip over the step edge 
increases for upward scans. \textbf{(c)} The tip-sample interaction 
potential $V(x_{\rm t})$ for the same three loads as in (b), illustrating 
the increase of barrier height and steepness with load. 
\textbf{(d)} Plot of the frictional increase as calculated in the 
Prandtl-Tomlinson model for the potential shown in (a). Parameters: 
$E_0 = 1.0$\,eV, $r_0 = 0.45$\,nm, $m_{\rm x} = m_{\rm z} = 10^{-10}$\,kg, 
$c_{\rm x} = 5.0$\,N/m and 
$\gamma_x = \gamma_z = 2 \sqrt{c_{\rm x} m_{\rm x}}$.}
\label{fig:simulation}
\end{figure}

The two-dimensional tip-sample interaction potential
$V_{\rm ts}(x_{\rm t},z_{\rm t})$ is computed by the summation of individual
Lenard-Jones potentials 
$ V_{\rm ts} = \sum_{i=1}^N E_0 \left( (r_0 / r_i)^{12}
		- 2 (r_0 / r_i)^6 \right)$ where $r_i$ represents the
distance between the point-like tip and the $i$th surface atom.
The parameters $E_0$ and $r_0$ describe the binding energy and the equilibrium
distance, respectively. Figure \ref{fig:simulation}(a) shows a color-coded
density plot of a tip-sample interaction potential calculated for a hexagonal
structure with an atomic distance of $a = 0.3$\,nm. By introducing this
potential into the equation of motion (\ref{eq:DGL}), we can compute the path
of the tip on the sample surface, obtaining the lateral force from
$F_{x} = c_{x} (x_{\rm M}-x_{\rm t})$.

Lateral force curves calculated in this way are plotted in
Fig.\,\ref{fig:simulation}(b) as a function of the loading force. Left and
right from the step edge, $F_{x}(x_{\rm M})$ exhibits for fixed loads the
typical saw tooth-like shape with the periodicity of the atomic lattice
expected from the classical PT model. This behavior occurs because the
condition
\begin{equation}
c_{\rm x} \leq \frac{\partial^2 ~ V_{\rm ts}}{{\partial x_{\rm t}}^2}
\label{eq:condition}
\end{equation}
is fulfilled \cite{Mate1987} and reproduces the tip behavior observed in 
actual FFM experiments on atomically flat terraces
\cite{Hoelscher2008a,Luethi1996a,Morita1996,Kerssemakers1995a,Hoelscher1998a,%
Hoelscher1999c,Socoliuc2004a}. The significantly altered tip movement in the
immediate vicinity of the surface step, however, requires a more careful
consideration.

For analysis, we start by computing the actual $z$-position of the tip from 
the stability condition
$F_{\rm load} = -\partial V_{\rm ts}/\partial z_{\rm t}$.
Calculating the tip-sample potential at these positions, we obtain
the potential the tip experiences for different loading forces
[Fig.\,\ref{fig:simulation}(c)]. Caused by a reduction of the atomic
coordination at the step edge, a step-induced potential barrier
(Schwoebel-Ehrlich barrier \cite{Ehrlich1966b,Schwoebel1966}) is obtained.
Conversely, increasing coordination leads to a potential minimum at the
bottom of the step. Left and right from the surface step, the potential shows
the sinusoidal shape assumed in the classical PT model.

The lateral force curves plotted in Fig.\,\ref{fig:simulation}(b) can now be
understood by imagining that the tip moves in this potential landscape during
scanning. For a downward scan [left to right in Fig.\,\ref{fig:simulation}(c)],
the tip needs three jumps to overcome the surface step [dashed arrows in
Figs.\,\ref{fig:PrandtlTomlinson} and \ref{fig:simulation}(c)]. First, it 
jumps over the Schwoebel-Ehrlich barrier into the minimum at the bottom of 
the step edge. The mechanism behind this movement is the same as for the 
earlier discussed stick-slip movement on the terraces: The tip sticks at the 
left slope of the maximum until the force of the spring is large enough to 
pull it over the barrier. The force needed for this process, however, will 
typically be larger than on the terraces. After this first jump, the tip is 
temporarily stuck in the minimum at the bottom of the step edge. A second 
larger-than-usual jump is needed to disengage the tip, which occurs again 
when the condition Eq.\,(\ref{eq:condition}) is fulfilled.
Finally, the tip jumps out of the shallow minimum nested within the rising
edge of the potential.

For an upward scan [right to left in Fig.\,\ref{fig:simulation}(c)], the tip
first makes a ``regular'' jump from the last minimum of the periodic potential 
of the terrace into a shallow minimum located within the potential's falling 
edge (first solid arrow). From there, it takes a smaller-than-usual jump into 
the well developed minimum at the bottom of the step edge (second solid arrow).
Pulling the tip now from this minimum over the barrier (third solid arrow) 
requires the application of a very large lateral force compared to typical 
forces experienced on the terraces. As a consequence,the tip subsequently 
jumps over a distance spanning several atomic unit cells, as can be seen 
from Fig.\,\ref{fig:simulation}(b).

From the analysis, it is evident that the frictional forces are markedly 
different for downwards and upwards scans. Because the tip has a much higher 
barrier as well as a much steeper rise (slope of the potential) to overcome 
during the upward scan compared to the downward scan, lateral forces are 
naturally much higher for scanning upwards. Also, we find that the relative 
height between the last minimum before the barrier and the maximum if 
approached from the side of the upper terrace does not significantly change 
with load, leading to a nearly load-independent frictional increase at the 
step edge. In contrast, both the relative barrier height as well as the slope 
increase if the barrier is approached from the side of the lower terrace, 
which leads to a linear dependence of the frictional increase on load. 
Fig.\,\ref{fig:simulation}(d) illustrates this behavior, where the frictional 
increase has been plotted for upwards and downward scans. 
Note that these results agree qualitatively very well with the experimental 
data of Figs.\,\ref{fig:HOPG} and \ref{fig:mos2}, suggesting that the 
experimentally observed load dependence is likely to be caused by the 
effects described in this paper. In this context, we would like to mention 
that similar results are obtained for a wide range of parameters as well as 
by representing the tip-sample interaction potential as a sum over Morse 
potentials.

In summary, we presented experiments and simulations analyzing the load 
dependence of atomic-scale friction at surface steps. Experimentally, a 
direction-dependence has been found, where the contribution of the 
frictional forces due to the presence of the step edge increases linearly 
with load for upwards scans while it is load-independent for downward scans. 
By introducing a modified Prandtl-Tomlinson model that includes an explicit 
description of the tip-sample interaction at surface steps, a theoretical 
basis for this behavior has been found. Finally, generalization of the above 
principles to other types of surface defects (vacancies, grain boundaries, 
etc.) where the atomic coordination will be temporarily altered suggests 
that this effect might very well dominate the macroscopic friction 
experienced on many materials.

It is a pleasure to thank Andr\'e Schirmeisen, Lars Jansen, Harald Fuchs
(University of M\"unster) and Volker Saile (Forschungszentrum Karlsruhe) for
discussions and their continuous support. This work was financially
supported by the BMBF (Grant No.\ 03N8704), the National Science Foundation
(Grant No.\ MRSEC DMR 0520495), and the Petroleum Research Fund of the
American Chemical Society (Grant No.\ PRF 42259-AC5).

%-------------------------------------------------------------------------------
% Produces the bibliography via BibTeX.
%\bibliography{SPMLiterature}
%-----------------------------------------------------------------------------

\end{document}